\documentclass[12pt]{article} 

\usepackage{graphics,color}
\usepackage{amssymb}
\usepackage{amsmath}
\usepackage{amsbsy}
\usepackage{epsfig}

\newcommand{\be}{\begin{equation}}
\newcommand{\ee}{\end{equation}}
\newcommand{\bea}{\begin{eqnarray}}
\newcommand{\eea}{\end{eqnarray}}
\newcommand{\bit}{\begin{itemize}}
\newcommand{\eit}{\end{itemize}}
\newcommand{\bc}{\begin{center}}
\newcommand{\ec}{\end{center}}
\newcommand{\bra}{\langle}
\newcommand{\ket}{\rangle}

\newcommand{\si}{\,\mbox{$\sum$}\hs{-0.47cm}\int}
\newcommand{\sgn}{\mbox{sgn}}
\newcommand{\im}{{\mathrm{Im}}}

\newcommand{\tr}{{\mathrm{tr}}}

\newcommand{\cE}{{\cal E}}
\newcommand{\cK}{{\cal K}}
\newcommand{\cM}{{\cal M}}
\newcommand{\om}{\omega}

\newcommand{\ko}{k^{0}}
\newcommand{\gv}{\boldsymbol\gamma}

\newcommand{\pv}{{\mathbf p}}
\newcommand{\kv}{{\mathbf k}}

\newcommand{\rv}{{\mathbf r}}
\newcommand{\xv}{{\mathbf x}}

\newcommand{\hm}{\hspace*{-0.6cm}}
\newcommand{\hs}[1]{\hspace*{#1}}

\newcommand{\half}{\frac{1}{2}}

\newcommand{\bean}{\begin{eqnarray*}}
\newcommand{\eean}{\end{eqnarray*}}
\newcommand{\nn}{\nonumber}
\newcommand{\veck}{{\mathbf k}}
\newcommand{\vecp}{{\mathbf p}}

\newcommand{\vecx}{{\mathbf x}}

\newcommand{\vecnul}{{\mathbf 0}}

\newcommand{\Kslash}{/\!\!\!\!K}

\newcommand{\Rslash}{/\!\!\!\!R}
\newcommand{\id}{{1\!\!1}}
\newcommand{\pbp}{{\bar p_+}}
\newcommand{\pbm}{{\bar p_-}}
\newcommand{\li}{{\rm Li}}

\begin{document}

\title{
\vskip -100pt
{\begin{normalsize}
\mbox{} \hfill SWAT 05-435\\
\vskip  30pt
\end{normalsize}}
{\bf\Large
Continuum and lattice meson spectral functions\\
at nonzero momentum and high temperature}
\author{
\addtocounter{footnote}{2}
Gert Aarts$^{a}$\thanks{email: g.aarts@swan.ac.uk} 
 {} and
Jose M.\ Mart{\'\i}nez Resco$^{b}$\thanks{email: martinezrescoj@brandonu.ca}
 \\ {} \\
{}$^a${\em\normalsize Department of Physics, University of Wales Swansea} 
\\
{\em\normalsize Singleton Park, Swansea, SA2 8PP, United Kingdom}
\\ {} \\
 {}$^b${\em\normalsize Department of Physics \& Astronomy, Brandon 
University} \\
   {\em\normalsize Brandon, Manitoba R7A 6A9, Canada }
}
}
\date{July 4, 2005}
\maketitle

\begin{abstract}
 We analyse discretization effects in the calculation of high-temperature 
meson spectral functions at nonzero momentum and fermion mass on the 
lattice. We do so by comparing continuum and lattice spectral functions in 
the infinite temperature limit. Complete analytical results for the 
spectral densities in the continuum are presented, along with simple 
expressions for spectral functions obtained with Wilson and staggered 
fermions on anisotropic lattices. We comment on the use of local and point 
split currents.

\end{abstract}

\newpage


\section{Introduction}

Motivated by the experimental progress in relativistic heavy ion 
collisions and the recreation of the quark gluon plasma, several questions 
have received substantial attention in the past few years. What happens to 
hadrons in the deconfined quark-gluon plasma? Do bound states persist? 
What is rate of photon and dilepton production from a hot QGP? How 
effectively are energy-momentum and charge transported? How long, or 
rather how short, are the typical relaxation times for hydrodynamic 
fluctuations?

Since this information is encoded in spectral functions, it is 
prohibitively difficult to access it directly from euclidean correlators 
obtained with lattice QCD, due to the intricacy of performing the 
analytical continuation from imaginary to real time. However, recent 
progress has been made by applying the Maximal Entropy Method (MEM) 
\cite{Asakawa:2000tr} to this problem. An (incomplete) list of high 
temperature studies includes the possible survival of hadronic bound 
states in the deconfined quark-gluon plasma 
\cite{Asakawa:2003re,Datta:2003ww,Umeda:2002vr}, thermal dilepton rates 
\cite{Karsch:2001uw}, and transport coefficients 
\cite{Gupta:2003zh}\footnote{We note here that Ref.\ 
\cite{Nakamura:2004sy} does not use an MEM analysis, but instead employs 
an Ansatz which was proposed in Ref.\ \cite{Karsch:1986cq} and criticized 
in Ref.\ \cite{Aarts:2002cc}.}.

In a spectral function investigation, the low-energy region $\om\lesssim 
T$ is of particular interest, since it is expected to be the most affected 
by nonperturbative medium effects. However, the reconstruction of spectral 
functions at small energies $\om\ll T$ is hindered by the insensitivity of 
euclidean correlators to details of spectral functions at these energies 
\cite{Aarts:2002cc}. This is especially important for the calculation of 
transport coefficients where by definition the interest is in the limiting 
value of current-current spectral densities as $\om\to 0$. Experience with 
the reconstruction of spectral densities in the low-energy region can be 
obtained by studying the simpler (but still nontrivial) problem of meson 
spectral functions at nonzero momentum above the deconfinement transition. 
Due to e.g.\ the scattering of quarks with gauge bosons below the 
lightcone (Landau damping), these spectral functions are expected to have 
a nontrivial structure. Since in the confined phase one expects to find 
mesons moving relative to the heatbath, described by simple quasiparticle 
spectral functions, increasing the temperature from below to above the 
transition temperature should result in a drastic change in those spectral 
functions.

Our aim in this paper is to provide a reference point for such an analysis 
on the lattice in the infinite temperature limit. It is therefore similar 
in spirit as Ref.\ \cite{Karsch:2003wy}, in which a study at zero momentum 
was performed. The paper is organized as follows. In the next section we 
give complete analytical expressions for continuum meson spectral 
functions at nonzero momentum and fermion mass in the infinite temperature 
limit and discuss several features. In Section \ref{lattice} we derive 
simple expressions for meson spectral functions for Wilson and staggered 
lattice fermions. We briefly comment on the value of the euclidean 
correlator at the midpoint and on the use of local and point split 
currents. The main results are shown in Section \ref{comparison}, where we 
contrast spectral functions obtained with Wilson and staggered fermions 
with the continuum results. Section \ref{summary} contains a short 
summary.


\section{Continuum}

We consider meson spectral functions with quantum numbers $H$, defined as
\be
\rho_H(t,\vecx) = \bra [ J_H(t,\vecx), J_H^\dagger(0,\vecnul)]\ket,
\ee
with $J_H(\tau,\xv) = \bar q(\tau,\xv)\Gamma_H q(\tau,\xv)$ and $\Gamma_H 
= \{\id,\gamma_5,\gamma^\mu, \gamma^\mu\gamma_5\}$.\footnote{In this 
section the gamma matrices obey $\gamma^{0\dagger}=\gamma^0$, 
$\gamma^{i\dagger}=-\gamma^i$, and $\gamma_5^\dagger=\gamma_5$. The 
anticommutation relations are $\{\gamma^\mu,\gamma^\nu\}=2g^{\mu\nu}$ and 
$\{\gamma^\mu,\gamma_5\}=0$ with $g^{\mu\nu}=$ 
diag$(+,-,-,-)$.} They are related to 
euclidean correlation functions, 
\be
\label{eqdef}
G_H(\tau,\xv) = \bra J_H(\tau,\xv) J_H^\dagger(0,\vecnul)\ket,
\ee
via the standard integral relation
\be
\label{eqGrho}
G_H(\tau,\pv) = \int_0^\infty \frac{d\om}{2\pi}\,
K(\tau,\om)\rho_H(\om,\pv),
\ee
with the kernel
\be
\label{eqkernel}
K(\tau,\om) = \frac{\cosh[\om(\tau-1/2T)]}{\sinh(\om/2T)}
= e^{\om\tau} n_B(\om) + e^{-\om\tau} \left[1+n_B(\om)\right],
\ee
where $n_B(\om)=1/(e^{\om/T}-1)$ is the Bose distribution.
At lowest order in the loop expansion, the euclidean 
correlators read in momentum 
space\footnote{$\gamma^0\Gamma_H^\dagger\gamma^0$ 
appears since the original correlator is of the form $\bra 
JJ^\dagger\ket$, not $\bra JJ\ket$.}
\be
\label{eqGH}
G_H(P) = -\si_K \tr\, S(K)\Gamma_H S(P+K)\gamma^0\Gamma_H^\dagger\gamma^0,
\ee
where $P=(i\om_n,\pv)$ with $\om_n=2\pi nT$ ($n\in\mathbb{Z}$) 
the Matsubara frequency in the imaginary-time formalism, and
\be
 \si_K  = T\sum_n \int_\veck, \;\;\;\;\;\;\;\;\;\;\;\;
 \int_\veck = \int \frac{d^3k}{(2\pi)^3}.
\ee
The fermion propagators are given by
\be
 \label{eqScont}
 S(K)=\frac{-1}{i\tilde{\om}_{n}\gamma^0 -\gv\cdot\kv -m}=
 -\int_{-\infty}^{\infty} 
 \frac{d\om}{2\pi}\frac{\rho_F(\om,\kv)}{i\tilde{\om}_{n}-\om},
\ee
where $\tilde{\om}_{n}=(2n+1)\pi T$ ($n\in \mathbb{Z}$) is a fermionic
Matsubara frequency and $\rho_F(\om,\kv)$ the spectral density of the 
fermion,
\be
 \rho_F(K) = \left(\Kslash+m\right)\rho(K)
 = \left(\Kslash+m\right)2\pi\,\sgn(\ko)\delta(k_0^2-\om_\kv^2),
\ee
with $\om_\kv = \sqrt{\kv^2+m^2}$.

Using the spectral representation for the fermion propagators, it is 
straightforward to arrive at
\bea
 \rho_H(P) =&&\hm 2\im\, G_H(i\om_n\to \om + i0^+,\vecp) \nn 
\\  \label{eqrhoH}
 = &&\hm N_c\int_{\kv,k^{0}} \tr\, 
 (\Kslash+m)\Gamma_H (\Rslash+m)\gamma^0\Gamma_H^\dagger\gamma^0 \,
 \rho(K)\rho(R) \left[ n_F(k^0) -n_F(r^0)\right],
\eea
with $P=(\om,\pv)$, $R=P+K$ and $n_F(\om)=1/(e^{\om/T}+1)$ is 
the Fermi distribution.

To facilitate the comparison with the lattice expressions below, we give 
here the result with the $k^0$ integral performed,
\bea
\rho_H(P) = &&\hm 2\pi N_c \int_\kv \bigg\{
\left(
a_H^{(1)} + a_H^{(2)}\frac{\kv\cdot\rv}{\om_\kv\om_\rv} 
+ a_H^{(3)}\frac{m^2}{\om_\kv\om_\rv} \right)
\left[ n_F(\om_\kv) - n_F(\om_\rv) \right] 
\delta(\om+\om_\kv-\om_\rv)
\nn \\ &&
+
\left(
a_H^{(1)} - a_H^{(2)}\frac{\kv\cdot\rv}{\om_\kv\om_\rv} 
- a_H^{(3)}\frac{m^2}{\om_\kv\om_\rv} \right)
\left[ 1 - n_F(\om_\kv) - n_F(\om_\rv) \right]
\delta(\om-\om_\kv-\om_\rv)
\nn \\ &&
-(\om\to-\om)
\bigg\}.
\label{eqrhocont}
\eea
The first line corresponds to scattering and contributes only below the 
lightcone ($\om^2<p^2$, Landau damping), while the second line corresponds 
to decay, contributing above threshold ($\om^2>p^2+4m^2$).
The coefficients $a_H^{(i)}$ arise from the three nonzero traces over the 
gamma matrices in Eq.\ (\ref{eqrhoH}) and depend on the channel 
under consideration. They are listed in Table \ref{table1}.

\begin{table}[t]
\begin{center}
\begin{tabular}{|c|c|c|c|c|c|}
\hline
& $\Gamma_H$ 	& $a_H^{(1)}$ & $a_H^{(2)}$ & $a_H^{(3)}$  \\
\hline
$\rho_{\rm S}$	&$\id$		& $1$ 	& $-1$	& $1$	\\
$\rho_{\rm PS}$ & $\gamma_5$	& $1$   & $-1$  & $-1$  \\
\hline
$\rho^{00}$	&$\gamma^0$	& $1$   & $1$   & $1$ 	\\
$\rho^{ii}$	&$\gamma^i$	& $3$   & $-1$ 	& $-3$ 	\\
$\rho_{\rm V}$ 	&$\gamma^\mu$  	& $2$  	& $-2$ 	& $-4$ 	\\
\hline
$\rho^{00}_5$	&$\gamma^0\gamma_5$	& $1$	& $1$	& $-1$ 	\\
$\rho^{ii}_5$	&$\gamma^i\gamma_5$	& $3$	& $-1$	& $3$ 	\\
$\rho_{\rm A}$	&$\gamma^\mu\gamma_5$	& $2$	& $-2$	& $4$ 	\\
\hline
\end{tabular}
 \caption{Coefficients $a_H^{(i)}$ for free spectral functions in 
 different channels $H$. In the case of $\gamma^i$ and $\gamma^i\gamma_5$, 
the sum is taken over $i=1,2,3$. We defined $\rho_{\rm V} = 
-g_{\mu\nu}\rho^{\mu\nu}$ and $\rho_{\rm A} = -g_{\mu\nu}\rho^{\mu\nu}_5$.
} 
\label{table1}
\end{center}
\end{table}

The remaining integrals can be performed as well. In terms of
\be
\bar p_\pm = \half\left[\om\pm p \beta(P)\right], 
\;\;\;\;\;\;\;\;
 \beta(P) = \sqrt{1-\frac{4m^2}{s}}, 
\;\;\;\;\;\;\;\;
s=\om^2-p^2,
\ee
the final expression in the continuum reads
\bea
 \nn 
 \rho_H(P) =&&\hm
 \Theta(s-4m^2)\frac{N_cT^2}{\pi} \bigg\{ 
\\ &&\hm \nn
  \frac{\beta(P)}{24T^2}\left[ 
  \left(3\om^2-p^2\beta^2(P)\right) a_H^{(1)} 
  + \left(3p^2-(3\om^2-2p^2)\beta^2(P)\right) a_H^{(2)}
  - 12m^2 a_H^{(3)}   
 \right]
 \\
 &&\hm \nn 
 +\frac{1}{4pT}  
\left[ 
 \left(\om^2-p^2\beta^2(P)\right) a_H^{(1)} 
 +\left(p^2-\om^2\beta^2(P)\right) a_H^{(2)}
  - 4m^2 a_H^{(3)} \right]
\ln \frac{1+e^{-\pbp/T}}{1+e^{-\pbm/T}}
\\ &&\hm \nn
 + \left( a_H^{(1)} + a_H^{(2)} \right)
\Big( \beta(P)
\left[\li_2(-e^{-\pbp/T})+\li_2(-e^{-\pbm/T})\right]
 \\ &&\hm \nn\hspace*{3cm}
 +\frac{2T}{p}\left[\li_3(-e^{-\pbp/T})-\li_3(-e^{-\pbm/T})\right] 
 \Big)
\bigg\} 
 \\ \nn
 &&\hm +\Theta(-s)\frac{N_cT^2}{\pi} \bigg\{
 \frac{1}{4pT}
\left[
 \left(\om^2-p^2\beta^2(P)\right) a_H^{(1)} 
 +\left(p^2-\om^2\beta^2(P)\right) a_H^{(2)}  - 4m^2 a_H^{(3)} \right]
\\ &&\hm \nn
  \ln \frac{1+e^{-\pbp/T}}{1+e^{\pbm/T}}
 + \left( a_H^{(1)} + a_H^{(2)} \right)
\Big( \beta(P)
\left[\li_2(-e^{-\pbp/T})-\li_2(-e^{\pbm/T})\right]
 \\ &&\hm \hspace*{5.5cm}
 +\frac{2T}{p}\left[\li_3(-e^{-\pbp/T})-\li_3(-e^{\pbm/T})\right] 
 \Big)
\bigg\}.
\label{eqrhoall}
\eea
We now discuss several features. First consider the asymptotic 
behaviour at large $\om$. We find that all 
spectral functions increase with $\om^2$,
\be
\label{eqasympt}
\lim_{\om\to \infty}
 \rho_H(P) =
 \Theta(s-4m^2)\frac{N_c}{8\pi}\om^2 
 \left( a_H^{(1)} - a_H^{(2)}\right),
\ee
as expected from naive dimensional arguments, except when there is a
cancellation. This happens for $\Gamma_H=\gamma^0, \gamma^0\gamma_5$, 
for which we find instead
\bea
 \lim_{\om\to\infty}\rho^{00}(P)
 = &&\hm \Theta(s-4m^2)\frac{N_c}{6\pi} p^2,  
\nn \\
\label{eqasympt2}
 \lim_{\om\to\infty}\rho^{00}_5(P) 
 =&&\hm
 \Theta(s-4m^2)\frac{N_c}{6\pi}\left(p^2+6m^2\right).
\eea
For the vector current this behaviour can be understood from current 
conservation $\partial_\mu j^\mu = 0$. Since at large $\om$ the effect of 
finite temperature is exponentially suppressed, we may use the zero 
temperature decomposition, 
\be
\rho^{\mu\nu}(P) = 2\im\, \Pi^{\mu\nu}_R(P) =  
2\left(P^2g^{\mu\nu}-P^\mu P^\nu\right) \im\, \Pi_R(P^2),
\ee
which explains the behaviour above. Current conservation also 
relates the other components of $\rho^{\mu\nu}$, 
\be
\om \rho^{00}(P) = p^i \rho^{i0}(P),
\;\;\;\;\;\;\;\;
\om \rho^{0j}(P) = p^i \rho^{ij}(P),
\;\;\;\;\;\;\;\;
\om^2\rho^{00}(P) = p^i p^j\rho^{ij}(P).
\ee
Since the axial vector current is not conserved, 
\be
\partial_\mu j^\mu_5 = 2m j_5 + 
\mbox{anomaly},
\ee
similar relations do not hold for $\rho_5^{\mu\nu}$. However, in the free 
case considered here, we find 
\be
\label{eqrho5}
\om^2\rho_5^{00}(\om,\vecnul) = 4m^2\rho_{\rm PS}(\om,\vecnul).
\ee
Any deviation from this is therefore due to the U(1)$_{\rm A}$ anomaly.

In the zero momentum limit, the spectral functions reduce to\footnote{A 
comparison with the coefficients $a_H$ and $b_H$ in 
Table 2.1 of Ref.\ \cite{Karsch:2003wy} yields 
$a_H^{(1)} - a_H^{(2)} = 2a_H$, $a_H^{(2)} - a_H^{(3)} = 2b_H$, except for 
the axial currents $(A_0, A_i)$, where we find $b_H=(1,-2)$ instead of 
$b_H=(0,3)$. Note that the coefficients in Ref.\ \cite{Karsch:2003wy} 
disagree with relation (\ref{eqrho5}).
Note also that the normalization differs by a factor of $2\pi$ and that 
the overall signs for $\rho_{\rm S}$ and $\rho_5$ are opposite.
}
\bea
 \rho_H(\om,\vecnul) =
 \Theta(\om^2-4m^2)\frac{N_c}{8\pi \om}
 \sqrt{\om^2-4m^2}\left[1-2n_F(\om/2)\right] 
 \Big[ \om^2\left( a_H^{(1)} -a_H^{(2)} \right) 
&&\hm \nn \\
+4m^2\left( a_H^{(2)} -a_H^{(3)} \right) \Big]
+2\pi \om\delta(\om) N_c
\left[ \left(a_H^{(1)} + a_H^{(2)}\right) I_1 +
\left( a_H^{(2)} -a_H^{(3)}\right) I_2 \right], 
&&\hm
\eea
with
\be
I_1 = -2\int_\kv n'_F(\om_\kv), 
\;\;\;\;\;\;\;\;\;\;\;\;
I_2 = -2\int_\kv \frac{k^2}{\om_\kv^2 }n'_F(\om_\kv).
\ee
In the massless case $I_1=I_2=T^{2}/6$. The term proportional to 
$\om\delta(\om)$ is all that remains from the scattering contribution 
below the lightcone in Eq.\ (\ref{eqrhocont}). It gives a $\tau$ 
independent contribution to the euclidean correlator since the kernel 
$K(\tau,\om) \sim 2T/\om$ for small $\om$. In particular, charge 
conservation dictates the form of $\rho^{00}$ and $G^{00}$ at zero momentum, 
\be  \label{eqG00}
 \rho^{00}(\om,\vecnul)=2\pi\chi\om\delta(\om),  \hs{2cm}
 G^{00}(\tau,\vecnul)=T\chi,
\ee
which is not altered by interactions, although the value of the charge 
susceptibility $\chi$ is. 
At the order computed here, $\chi=2N_{c}I_{1}$.
For completeness we give here the euclidean correlator at zero momentum 
and mass
\be
G_H(\tau,\vecnul) = \frac{N_cT^3}{6} \left[
 a_H^{(1)}+a_H^{(2)}
+\frac{3}{2} \left(a_H^{(1)}-a_H^{(2)}\right)
\frac{3u+u\cos(2u)-2\sin(2u)}{\sin^3u}
\right],
\ee
where $u=2\pi T (\tau -1/2T)$.

Finally, it follows from the spectral decomposition
\be
\label{eqspecdec}
\rho_H(P) = \frac{1}{Z}\sum_{n,m} |\bra n|J_H(0)|m\ket|^2 
(2\pi)^4\delta^4(P+P_n-P_m) \left(e^{-p^0_n/T} -e^{-p^0_m/T}\right),
\ee
where $Z$ is the partition function, that all spectral functions for a 
single current $J_H$ are odd and positive semi-definite for positive 
argument, i.e.\ $\om\rho_H(\om,\pv)\geq 0$. Obviously, spectral 
functions that are defined as the difference between such spectral 
functions, such as $\rho_{\rm V} = \rho^{ii}-\rho^{00}$ and $\rho_{\rm A} 
= \rho^{ii}_{5}-\rho^{00}_{5}$ can turn negative. Indeed, it is easy to see 
that $\rho_{\rm V}(\om,\pv)$ is 
negative for small $\om$ if $p^2<2m^2$. All other spectral functions 
increase linearly with $\om$ for small $\om$ and nonzero $\pv$.

Although not the topic of this paper, we briefly mention how corrections 
due to interactions appear at very high temperature. First of all, for 
soft momentum $|\pv| \sim \om \sim gT$, a hard thermal loop 
\cite{Braaten:1989mz} calculation is needed, see e.g.\ Refs.\ 
\cite{Braaten:1990wp,Karsch:2000gi,Alberico:2004we} for such studies. The 
gap in the spectrum for $p^2<\om^2<p^2+4m^2$ is filled when two loop 
diagrams are included, due to e.g.\ bremsstrahlung \cite{Aurenche:1998nw}. 
Around the lightcone the loop expansion breaks down due to the 
Landau-Pomeranchuk-Migdal effect and an infinite series of ladder diagrams 
contribute at leading order in the strong coupling constant
\cite{Arnold:2001ms,Aurenche:2002wq}.
 Finally for very soft momenta and energies, the structure of 
current-current spectral functions is determined by general hydrodynamical 
considerations \cite{Kadanoff}. So far a diagrammatic calculation in this 
regime has been carried out only in the case of the spatial vector 
spectral function $\rho^{ii}(\om,\vecnul)$ in the limit of exactly zero 
momentum and vanishing energy $\om\to 0$, which is relevant for the 
electrical conductivity: see 
Refs.\ \cite{ValleBasagoiti:2002ir,Aarts:2002tn} for 
details on the weak coupling result at leading-logarithmic order and 
Ref.\ \cite{Aarts:2005vc} for the large $N_f$ result.


\section{Lattice}
\label{lattice}


\subsection{Wilson fermions}

In this section we derive expressions for meson spectral functions on a 
lattice with $N_\sigma^3\times N_\tau$ sites. The lattice spacing is 
denoted with $a$ in the spatial directions and with $a_\tau$ in the 
temporal direction, $\xi=a/a_\tau$ is the anisotropy parameter. The 
temperature is related to the extent in the imaginary time direction, $T = 
1/(N_\tau a_\tau)$. We start with standard Wilson fermions. The lattice 
fermion 
propagator (with coefficients $r_4, r_{\rm space}=r$) reads\footnote{In 
this section the gamma matrices are hermitian, 
$\gamma_\mu^\dagger=\gamma_\mu$, $\gamma_5^\dagger=\gamma_5$, and obey 
$\{\gamma_\mu,\gamma_\nu\} = 2\delta_{\mu\nu}$, $\{\gamma_\mu,\gamma_5\} = 
0$. They are related to the gamma matrices of the previous section as 
$\gamma_4=\gamma^0$, $\gamma_i = -i\gamma^i$. We use lattice units $a=1$.}
 \be
S(K) = \frac{-i\gamma_4\sin k_4 -i \cK_\kv + r_4\left(1-\cos k_4\right) 
+\cM_\kv}{\sin^2 k_4 +\cK^2_\kv + \left[ r_4\left(1-\cos 
k_4\right)+\cM_\kv\right]^2},
\ee
where
\be
\cK_\kv = \frac{1}{\xi}\sum_{i=1}^3 \gamma_i\sin k_i, 
\;\;\;\;\;\;\;\;\;\;\;\;
\cM_\kv = \frac{1}{\xi}\left[ r \sum_{i=1}^3 \left(1-\cos 
k_i\right) + m\right].
\ee 
 We use periodic boundary conditions in space, $k_i = 2\pi n_i/N_\sigma$ 
with $n_i = -N_\sigma/2+1, -N_\sigma/2+2, \ldots, N_\sigma/2-1, 
N_\sigma/2$ for $i=1, 2, 3$, and antiperiodic boundary conditions in 
imaginary time, $k_4 = \pi(2n_4+1)/N_\tau$ with $n_4 = -N_\tau/2+1, 
-N_\tau/2+2, \ldots, N_\tau/2-1, N_\tau/2$.

To make a smooth connection with the expressions in the continuum we 
follow Ref.\ \cite{Karsch:2003wy} and use the mixed representation of 
Carpenter and Baillie \cite{Carpenter:1984dd}
\be \label{eqSCB}
 S(\tau, \kv) = \gamma_4 S_4(\tau, \kv) + 
 \sum_{i=1}^3 \gamma_i S_i(\tau, \kv) +\id S_u(\tau, \kv).
\ee
In order to avoid the doubler in the time direction, we proceed with 
$r_4=1$, so that (for $0\leq \tau<1/T$)
\bea
\nn
S_4(\tau, \kv) =&&\hm S_4(\kv)\cosh(\tilde\tau E_\kv), \\
\nn
S_i(\tau, \kv) =&&\hm S_i(\kv)\sinh(\tilde\tau E_\kv), \\
S_u(\tau, \kv) =&&\hm S_u(\kv)\sinh(\tilde\tau E_\kv)
- \frac{\delta_{\tau0}}{2(1+\cM_\kv)}.
\eea
Here $\tilde\tau = \tau-1/2T$ and 
\bea
\nn
S_4(\kv) =&&\hm \frac{\sinh \left(E_\kv/\xi\right)}{2\cE_\kv\cosh(E_\kv/2T)}, \\
\nn
S_i(\kv) =&&\hm \frac{1}{\xi} \frac{i\sin k_i}{2\cE_\kv\cosh(E_\kv/2T)}, \\
S_u(\kv) =&&\hm 
-\frac{1-\cosh \left(E_\kv/\xi\right) +\cM_\kv}{2\cE_\kv\cosh(E_\kv/2T)},
\label{eqsss}
\eea
with $\cE_\kv = (1+\cM_\kv) \sinh \left(E_\kv/\xi\right)$.
The single particle energy $E_\kv$ is determined by\footnote{The factor 
$1/\xi$ is included so that in the continuum limit $E_\kv\to 
\om_\kv=\sqrt{\kv^2+m^2}$ (with $a=1$).}
 \be
\cosh \left(E_\kv/\xi\right) = 1+ 
\frac{\cK_\kv^2+\cM_\kv^2}{2(1+\cM_\kv)}.
\ee
 The final term in $S_u(\tau,\kv)$ is the sole remnant of the 
nonpropagating time doubler; below we consider $0<\tau<1/T$. The 
propagator satisfies $S(-\tau,\kv) = \gamma_5 
S^\dagger(\tau,\kv)\gamma_5$.

The correlators we are interested in are of the form
\be
G_H(\tau,\pv) = -\frac{N_c}{L^3} \sum_\kv \tr\, S(\tau,\kv) \Gamma_H 
S(-\tau,\rv) \Gamma_H,
\ee
where again $\rv=\pv+\kv$.
Inserting Eq.\ (\ref{eqSCB}) gives the euclidean correlator\footnote{Note that we 
now start from $\bra J_H(\tau,\vecx) J_H(0,\vecnul)\ket$ rather than from Eq.\ 
(\ref{eqdef}). This only affects the overall sign in some channels, which has 
been adjusted to agree with the continuum one.}
\bea
\nn
G_H(\tau,\pv) =  \frac{4N_c}{L^3} \sum_\kv \Big[
a_H^{(1)} S_4(\tau,\kv)S_4^\dagger(\tau,\rv) 
- a_H^{(2)} \sum_i S_i(\tau,\kv)S_i^\dagger(\tau,\rv) 
&&\hm\\
- a_H^{(3)} S_u(\tau,\kv)S_u^\dagger(\tau,\rv) \Big], &&\hm
\label{eqG}
\eea
where the coefficients $a_H^{(i)}$ are the same as before (see Table 
\ref{table1}). 

We will now extract the spectral functions in a form that closely 
resembles the continuum expressions. 
In the terms $SS^\dagger$ we encounter products of hyperbolic 
functions. These can be written as
\bea
\nn
\sinh(\tilde\tau E_\kv) \sinh(\tilde\tau E_\rv) =
\frac{1}{4} \int_{-\infty}^\infty d\om \cosh (\om\tilde\tau)
\Big[
\delta(\om -E_\kv-E_\rv) + \delta(\om +E_\kv+E_\rv) &&\hm 
\\ 
-\delta(\om -E_\kv+E_\rv) - \delta(\om +E_\kv-E_\rv) \Big], &&\hm 
\eea
and similarly for the product of two hyperbolic cosines. 
Noting that the factor $\cosh(\om\tilde\tau)$ is the sole place 
with $\tau$ dependence and that it is of the same form as in the 
kernel (\ref{eqkernel}), it is straightforward to write the above 
expression for $G_H(\tau,\pv)$ as
\be
G_H(\tau,\pv) =  \int_{0}^\infty \frac{d\om}{2\pi}\, 
K(\tau,\om) \rho^{\rm Wilson}_{H}(\om,\pv),
\ee
and read off the expressions for the lattice spectral functions,
\bea
&&\hm
\rho^{\rm Wilson}_H(P) = 
\frac{4\pi N_c}{L^3}\sum_\kv \sinh\left(\frac{\om}{2T}\right) 
\bigg\{ 
\nn \\ &&\hm \;\;\;\;\;\;\;\;
\bigg[ 
a_H^{(1)} S_4(\kv)S^\dagger_4(\rv) 
+ a_H^{(2)} \sum_i S_i(\kv)S^\dagger_i(\rv)
+ a_H^{(3)} S_u(\kv)S^\dagger_u(\rv) \bigg] 
\delta(\om +E_\kv-E_\rv)
\nn \\&&\hm \;\;\;\;\;\;\;\;
+ \bigg[
a_H^{(1)} S_4(\kv)S^\dagger_4(\rv) 
- a_H^{(2)} \sum_i S_i(\kv)S^\dagger_i(\rv)
- a_H^{(3)} S_u(\kv)S^\dagger_u(\rv) \bigg]
\delta(\om -E_\kv-E_\rv)
\nn \\ &&\hm \;\;\;\;\;\;\;\;
+ (\om\to-\om)
\bigg\}.
\label{eqrhoW}
\eea
This result can be directly compared with the continuum expression 
(\ref{eqrhocont}), using Eq.\ (\ref{eqsss}) and realizing that
\be
\frac{\sinh(\om/2T)}{\cosh(E_\kv/2T)\cosh(E_\rv/2T)} = 
\begin{cases}
2\left[ n_F(E_\kv) -  n_F(E_\rv) \right] & 
\;\;{\rm if}\;\; \om=E_\rv-E_\kv, 
\\
2\left[ 1 - n_F(E_\kv) - n_F(E_\rv) \right] & 
\;\; {\rm if}\;\; \om=E_\rv+E_\kv.
\end{cases}
\ee


\subsection{Staggered fermions}

In the case of staggered fermions we perform the analysis with 
naive fermions, since this leads to equivalent results 
\cite{Kawamoto:1981hw}. Taking therefore $r_4=r_{\rm space}=0$ 
yields the fermion propagator (\ref{eqSCB}) with 
\bea
\nn
S_4(\tau, \kv) =&&\hm S_4(\kv) \left(1-(-1)^{\tau/a_\tau}\right) 
\cosh(\tilde\tau E_\kv), \\
\nn
S_i(\tau, \kv) =&&\hm S_i(\kv) \left(1+(-1)^{\tau/a_\tau}\right) 
\sinh(\tilde\tau E_\kv), \\
S_u(\tau, \kv) =&&\hm S_u(\kv) \left(1+(-1)^{\tau/a_\tau}\right) 
\sinh(\tilde\tau E_\kv),
\label{eqSstag}
\eea
where $S_4(\kv)$ and $S_i(\kv)$ are as in Eq.\ (\ref{eqsss}) with
$\cE_\kv = \cosh \left(E_\kv/\xi\right)\sinh \left(E_\kv/\xi\right)$, and 
\be
S_u(\tau, \kv) = - 
\frac{1}{\xi} \frac{m}{2\cE_\kv\cosh(E_\kv/2T)}.
\ee
The single particle energy $E_\kv$ is now determined by 
\be
\cosh \left(E_\kv/\xi\right) = \sqrt{ 1 + \cK_\kv^2+(m/\xi)^2}.
\ee
Using the same steps as before, the euclidean meson correlator takes again 
the form (\ref{eqG}) and can be written in a spectral representation as
\be
G_H(\tau,\pv) = 2\int_0^\infty \frac{d\om}{2\pi} K(\tau,\om) \left[ 
\rho_H^{\rm naive}(\om,\pv) - (-1)^{\tau/a_\tau} 
\tilde\rho_H^{\rm naive}(\om,\pv)\right],
\ee
with the same kernel as above. 

The desired spectral function $\rho_H^{\rm naive}(\om,\pv)$ is exactly as 
in Eq.\ (\ref{eqrhoW}), whereas the staggered partner $\tilde\rho_H^{\rm 
naive}(\om,\pv)$ has the same form but with coefficients $\tilde a_H^{(1)} 
= a_H^{(1)}$, $\tilde a_H^{(2)} = -a_H^{(2)}$, $\tilde a_H^{(3)} = 
-a_H^{(3)}$. This staggered contribution $\tilde\rho_H$ represents the 
spectral function in the channel related to the original $\rho_H$ by 
replacing $\Gamma_H \to \tilde\Gamma_H=\gamma_4\gamma_5\Gamma_H$ 
\cite{Karsten:1980wd}. Note that in particular the pseudoscalar (scalar) 
spectral function mixes with the zero'th component of vector (axial 
vector) current spectral function. In an actual MEM investigation, the 
staggered partners can be disentangled using an independent analysis on 
even/odd timeslices, which yields the linear combinations $\rho_H^{\rm 
naive}(\om,\pv) \mp \tilde \rho_H^{\rm naive}(\om,\pv)$. Finally, in order 
to compare the naive lattice spectral functions with the continuum and the 
Wilson ones, we divide $\rho_H^{\rm naive}$ by a factor of 8, which takes 
care of the space doublers.


\subsection{Midpoint of the euclidean correlator}
\label{midpoint}

In the midpoint $\tau=1/2T$ ($\tilde\tau=0$), the hyperbolic functions 
in the fermion propagator $S(\tau, \pv)$ take simple values, and it is 
easy to see that
\be
 G_H(1/2T,\pv)=\frac{4N_c}{L^3}\sum_\kv a^{(1)}_H S_4(\kv)S_4^\dagger(\rv).
\ee
This implies that the channel dependence of the value at the midpoint 
enters only via $a^{(1)}_H$.

Analogous expressions hold for naive fermions and in the continuum, so 
that one can write
\be
G_H(1/2T,\pv) = a_H^{(1)} C(\pv),
\ee
with
\bea
C_{\rm continuum}(\pv) = &&\hm
N_c\int_\kv \frac{1}{\cosh(\om_\kv/2T)}\frac{1}{\cosh(\om_\rv/2T)},
\nn \\
C_{\rm Wilson}(\pv) = &&\hm
\frac{N_c}{L^3} \sum_\kv
\frac{1}{(1+\cM_\kv)\cosh(E_\kv/2T)}\frac{1}{(1+\cM_\rv)\cosh(E_\rv/2T)},
\nn \\
C_{\rm naive}^{N_\tau/2 \,\,\rm odd}(\pv) = &&\hm
\frac{4N_c}{L^3} \sum_\kv
\frac{1}{\cosh(E_\kv/\xi)\cosh(E_\kv/2T)}
\frac{1}{\cosh(E_\rv/\xi)\cosh(E_\rv/2T)},
\nn \\
C_{\rm naive}^{N_\tau/2 \,\,\rm even}(\pv) = &&\hm 0.
\eea
Combining this result with the relation between the euclidean
correlator and the spectral function $\rho_H$ in Eq.\
(\ref{eqGrho}), yields a constraint for the free spectral density
\be  \label{eqsum1}
 G_H(1/2T,\pv)=\int_0^\infty \frac{d\om}{2\pi} 
 \frac{\rho_H(\om,\pv)}{\sinh(\om/2T)}=a_H^{(1)} C(\pv).
\ee
In the case of naive fermions this gives
\be
 G_H^{\rm naive}(1/2T,\pv) = 2\int_0^\infty \frac{d\om}{2\pi}
 \frac{1}{\sinh(\om/2T)} \left[\rho_H^{\rm naive}(\om,\pv)\mp
 \tilde\rho_H^{\rm naive}(\om,\pv)\right],
\ee
for $N_\tau/2$ even/odd, from which we find
\be  \label{eqsum2}
 \int_0^\infty \frac{d\om}{2\pi}
 \frac{\rho_H^{\rm naive}(\om,\pv)}{\sinh(\om/2T)}
=
 \int_0^\infty \frac{d\om}{2\pi}
 \frac{\tilde\rho_H^{\rm naive}(\om,\pv)}{\sinh(\om/2T)}
 =\frac{1}{4} a_H^{(1)} C_{\rm naive}^{N_\tau/2\,\,\rm odd}(\pv).
\ee 
 Although the free spectral functions in the various channels are 
distinctly different, we conclude that the integral of 
$\rho_H(\om,\pv)/\sinh(\om/2T)$ is in all cases related to $C(\pv)$ given 
above, both in the continuum and on the lattice.


\subsection{Point split current}

The local vector current we have considered so far, 
$j_\mu=\bar\psi\gamma_\mu\psi$, is not exactly conserved on the lattice. 
Instead, the conserved current is
\be
j_\mu(x) = \bar\psi(x+\hat a_\mu)P_\mu^+\psi(x) - 
\bar\psi(x)P_\mu^-\psi(x+\hat a_\mu),
\ee
with $P_\mu^\pm = \half(r_\mu\pm\gamma_\mu)$, $r_i\equiv r$ and 
$\hat a_\mu = \hat\mu a_\mu$. Here we present 
a short analysis comparing the two currents.

A correlator especially sensitive to the difference between the local and 
the conserved current is $G^{00}(\tau,\vecnul)$ at vanishing momentum, 
since its form is determined by charge conservation, see Eq.\ 
(\ref{eqG00}). In particular it should be $\tau$ independent. 
On the lattice, the correlator for the local current is
\be
G^{00}(\tau,\vecnul) =  \frac{4N_c}{L^3} \sum_\kv \Big[
 \left| S_4(\tau,\kv)\right|^2
-  \sum_i \left|S_i(\tau,\kv)\right|^2
- \left| S_u(\tau,\kv)\right|^2
\Big].
\ee
After some algebra, this can be written as
\bea
\nn
G^{00}_{\rm Wilson}(\tau,\vecnul) =&&\hm  \frac{N_c}{L^3}\sum_\kv 
\frac{1}{(1+\cM_\kv)^2\cosh^2(E_\kv/2T)}, \\
G^{00}_{\rm naive}(\tau,\vecnul) = &&\hm \frac{2N_c}{L^3}\sum_\kv 
\frac{1-(-1)^{\tau/a_\tau}\cosh(2\tilde \tau E_\kv)}
{\cosh^2(E_\kv/\xi)\cosh^2(E_\kv/2T)}.
\eea
 For Wilson fermions we find a $\tau$ independent result at leading order. 
However, since the local current is not related to a symmetry, dependence 
on $\tau$ is expected to arise when interactions are present. This is easy 
to study in actual simulations. For naive fermions we indeed find a $\tau$ 
dependent result.

With the conserved current the situation should be different. We find
for Wilson fermions, with $r_4=1$, 
\bea
&&\hm G^{00}_{\rm Wilson}(\tau,\vecnul) = \frac{4N_c}{L^3}\sum_\kv \Big\{
S_4(\tau-a_\tau,\kv)S_4^\dagger(\tau+a_\tau,\kv)
-S_u(\tau-a_\tau,\kv)S_u^\dagger(\tau+a_\tau,\kv)
\nn \\ 
&&\hm 
-\sum_i |S_i(\tau,\kv)|^2
-S_4(\tau-a_\tau,\kv)S_u^\dagger(\tau+a_\tau,\kv)
+S_u(\tau-a_\tau,\kv)S_4^\dagger(\tau+a_\tau,\kv)
\Big\}, 
\eea
and for naive fermions
\bea
&&\hm
G^{00}_{\rm naive}(\tau,\vecnul) = \frac{2N_c}{L^3}\sum_\kv \Big\{
S_4(\tau-a_\tau,\kv)S_4^\dagger(\tau+a_\tau,\kv)
+ |S_4(\tau,\kv)|^2 - |S_u(\tau,\kv)|^2 
\nn \\ &&\hm
-S_u(\tau-a_\tau,\kv)S_u^\dagger(\tau+a_\tau,\kv)
-\sum_i \left[ S_i(\tau-a_\tau,\kv)S_i^\dagger(\tau+a_\tau,\kv)
+ |S_i(\tau,\kv)|^2 \right]
\Big\}. \;\;\;\;
\eea
Indeed, this yields the anticipated result for a conserved current,
\be
G^{00}_{\rm Wilson} (\tau,\vecnul) = \half G^{00}_{\rm naive}(\tau,\vecnul) 
= \frac{N_c}{L^3}\sum_\kv \frac{1}{\cosh^2(E_\kv/2T)}
= -\frac{N_c}{L^3}\sum_\kv 4Tn_F'(E_\kv).
\ee
 In both cases $G^{00}(\tau,\vecnul)$ is now $\tau$ independent; this 
should remain to be the case when interactions are included. Moreover, the 
lattice susceptibility takes the same form as in the continuum, see 
below Eq.\ (\ref{eqG00}). The factor $1/2$ in the naive case appears 
because of the contribution from the time doublers.

If one is interested in the reconstruction of vector spectral functions 
for e.g.\ thermal dilepton production \cite{Karsch:2001uw}, it may be 
important to use the properly conserved current. It would therefore be 
interesting to compare spectral functions obtained with local and point 
split currents in the interacting case.


\section{Comparison}
\label{comparison}

\begin{figure}[ht]
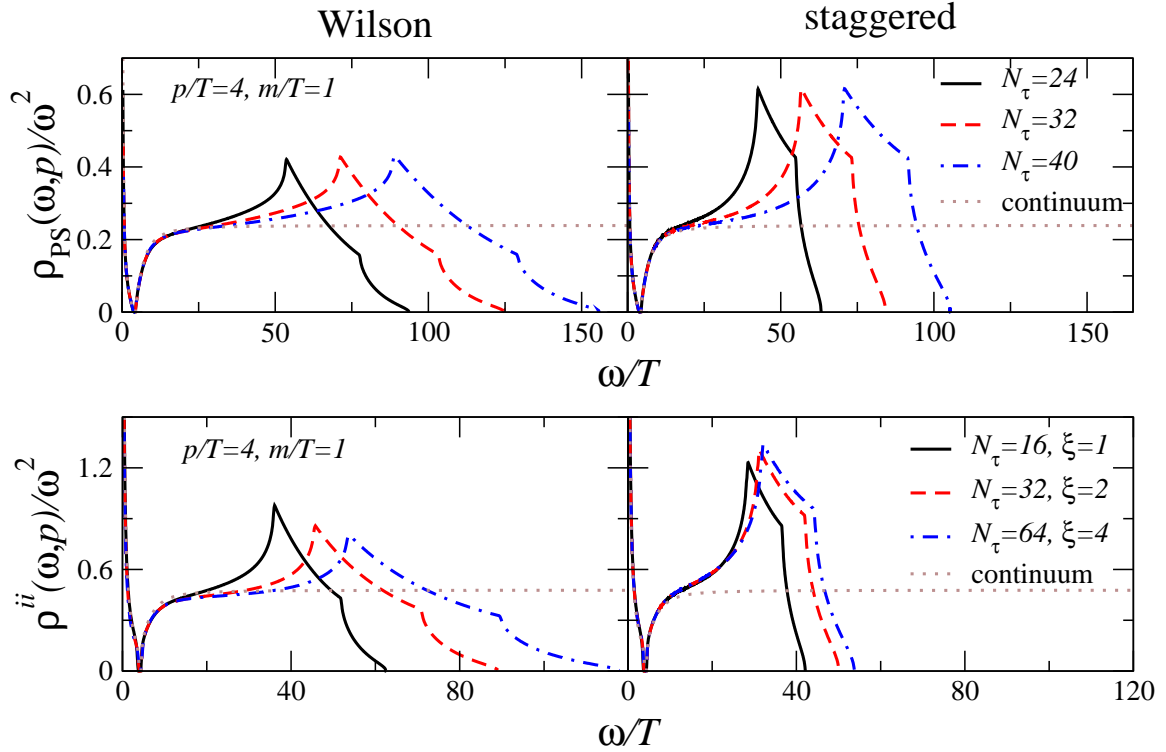

\centerline{\epsfig{figure=plotPSNt.eps,width=15cm}}
\vspace*{0.3cm}
\centerline{\,\,\,\,\,\epsfig{figure=plotiiani2.eps,width=15.2cm}}
\caption{Pseudoscalar spectral functions $\rho_{\rm
PS}(\om,\pv)/\om^2$ as a function of $\om/T$ for $N_\tau=24, 32, 40$
and $\xi=1$ (above) and vector spectral functions
$\rho^{ii}(\om,\pv)/\om^2$ as a function of $\om/T$ for $(N_\tau,
\xi)=(16,1), (32, 2), (64, 4)$ (below). In both cases $p/T=4, m/T=1$.
}
\label{figplotPSNt}
\end{figure}

We now contrast the meson spectral functions obtained 
for free Wilson and staggered lattice fermions with the continuum ones. 
The lattice meson spectral functions obtained above can be analysed for 
finite $N_\sigma$ and $N_\tau$. For small $N_\sigma$ the discreteness 
inherent in the definition of spectral functions (see e.g.\ the spectral 
decomposition (\ref{eqspecdec})) is clearly visible. Following Ref.\ 
\cite{Karsch:2003wy} we therefore take the thermodynamic limit 
$N_\sigma\to \infty$ and focus on the effect of finite 
$N_\tau$.\footnote{In practice we take $N_\sigma\sim 1500-2000$, replace 
the delta functions in Eq.\ (\ref{eqrhoW}) with block functions with width 
$\Delta \om$ and height $1/\Delta \om$, and divide the $\om$ interval in 
$N_\om$ bins. We used $N_\om=1000$. The bin width is determined by $\Delta 
\om=\om_{\rm max}/N_\om$ where $\om_{\rm max}$ is discussed below. See 
also \cite{Karsch:2003wy}.} In all figures the nonzero external momentum 
$p=4T$ and the fermion mass $m=T$. For Wilson fermions we show results 
with $r=1$. The anisotropy parameter $\xi=1$, except in the bottom part of 
Fig.\ \ref{figplotPSNt}. We only show meson spectral functions 
obtained with local operators.

\begin{figure}[t]
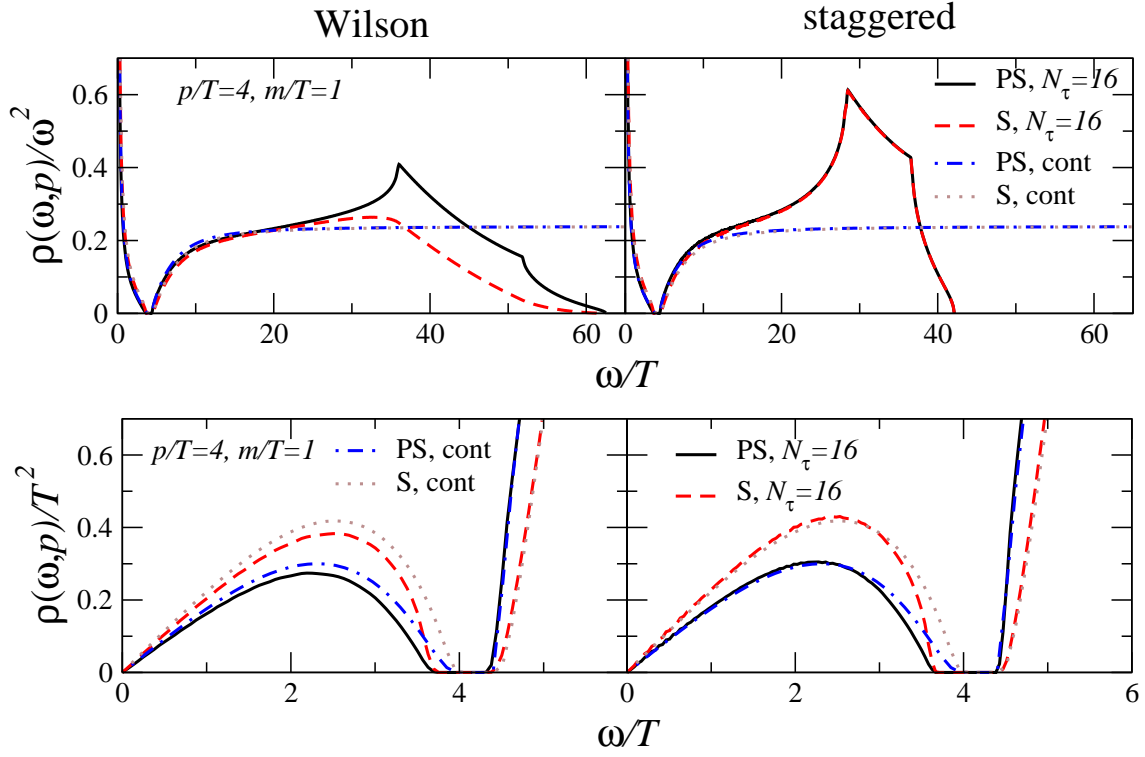

\centerline{\epsfig{figure=plotPSS.eps,width=15cm}}
\vspace*{0.3cm}
\centerline{\,\,\epsfig{figure=plotPSSzoom2.eps,width=15cm}}
\caption{Pseudoscalar and scalar spectral functions
$\rho_{\rm PS,S}(\om,\pv)/\om^2$ (above) and
$\rho_{\rm PS,S}(\om,\pv)/T^2$ (below)
as a function of $\om/T$ for $N_\tau=16$, $p/T=4, m/T=1$ and $\xi=1$.
}
\label{figplotPSS}
\end{figure}

In Fig.\ \ref{figplotPSS} we show the scalar and pseudoscalar spectral
functions for Wilson (left) and staggered (right) fermions. In order to
emphasize the effects of the lattice cutoff, $\rho_{\rm PS,S}$ is
divided by $\om^2$ in the top figures. The continuum result then reaches a
constant value ($3/4\pi$) for large $\om$, see Eq.\ (\ref{eqasympt}).
Instead, on the lattice there is a maximal energy $\om_{\rm max}$,
determined by the delta function $\delta(\om-E_\kv-E_\rv)$. Since the
external momentum $\pv$ is small with respect to momenta at the edge of
the Brillouin zone, the maximum value for Wilson fermions (with $r=1$) is
determined by fermion momenta $\kv=(\pi/a,\pi/a,\pi/a)$
\cite{Karsch:2003wy}, which gives
\be
\label{ommaxW}
\frac{\om_{\rm max}^{\rm Wilson}}{T} \approx
2N_\tau\ln\left(1+\frac{6+am}{\xi}\right),
\ee
and for staggered fermions by fermion momenta 
$\kv=(\pi/2a,\pi/2a,\pi/2a)$, which yields
\be
\label{ommaxS}
\frac{\om_{\rm max}^{\rm naive}}{T} \approx 2N_\tau\ln
\frac{\sqrt{\xi^2+3+a^2m^2}+\sqrt{3+a^2m^2}}{\xi}.
\ee
 The maximum value is smaller for staggered fermions. The cusps in the 
plots originate from the corners of the Brillouin zone. Both for continuum 
and staggered fermions, we find that the scalar and pseudoscalar channel 
are indistinguishable for large $\om$. The reason is that the finite 
fermion mass is negligible for such large energies. In the case of Wilson 
fermions the Wilson mass term breaks the chiral symmetry completely and 
the scalar and pseudoscalar spectral functions differ.
                                                                                
\begin{figure}[t]
\centerline{\epsfig{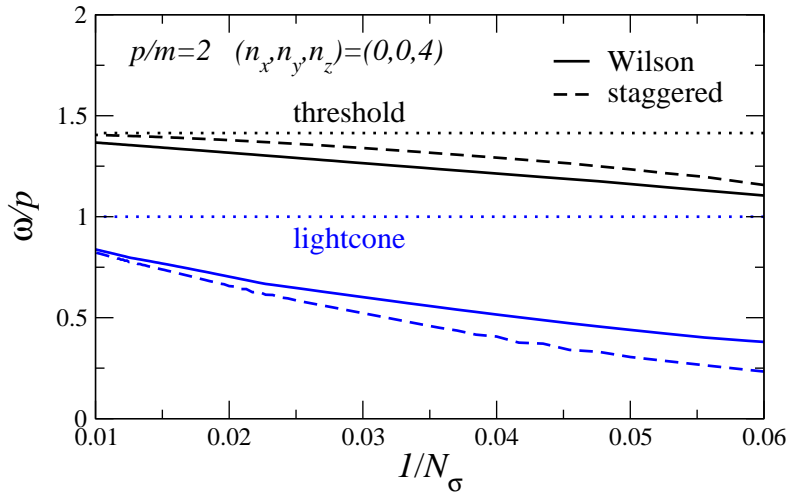}}
\caption{Effect of finite lattice spacing on the 'lightcone' and 
'threshold' for fixed $pL\approx 25$ and $m=p/2$. The continuum values are 
$\om/p=1$ resp.\ $\om/p=\sqrt{1+4m^2/p^2}=\sqrt{2}$.
}
\label{figlc_thr}
\end{figure}
                                                                                
The spectral functions vanish for energies $p<\om<\sqrt{p^2+4m^2}$. The 
physically interesting contribution below the lightcone appears as a 
divergent one in the top plots. We therefore show $\rho_{\rm PS,S}/T^2$ in 
the plots on the bottom. The spectral functions increase linearly for 
small $\om$ and vanish at the lightcone. Due the finite fermion mass, the 
scalar and the pseudoscalar channel are now physically distinct. The main 
lattice artefact in this region appears to be the mismatch between the 
location of the lightcone in the continuum and the lattice theory. This is 
due to the difference between continuum and lattice dispersion relations. 
To study this further, we define the lattice 'lightcone' and 'threshold' 
via
 \be
\mbox{lightcone:} \;\;\;\om = \max_\kv(E_\kv-E_{\kv+\pv}),
\;\;\;\;\;\;\;\;\;
\mbox{threshold:} \;\;\;\om = \min_\kv(E_\kv+E_{\kv+\pv}).
\ee
In Fig.\ \ref{figlc_thr} we show the result as a function of
$1/N_\sigma\sim a$ for fixed momentum $\pv=(0,0, 8\pi/aN_\sigma)$ 
($pL\approx 25$) and $m=p/2$.
As expected, the continuum and lattice results agree for decreasing
$1/N_\sigma$ (decreasing lattice spacing), but for finite $a$ the
corrections can be substantial.

\begin{figure}[t]
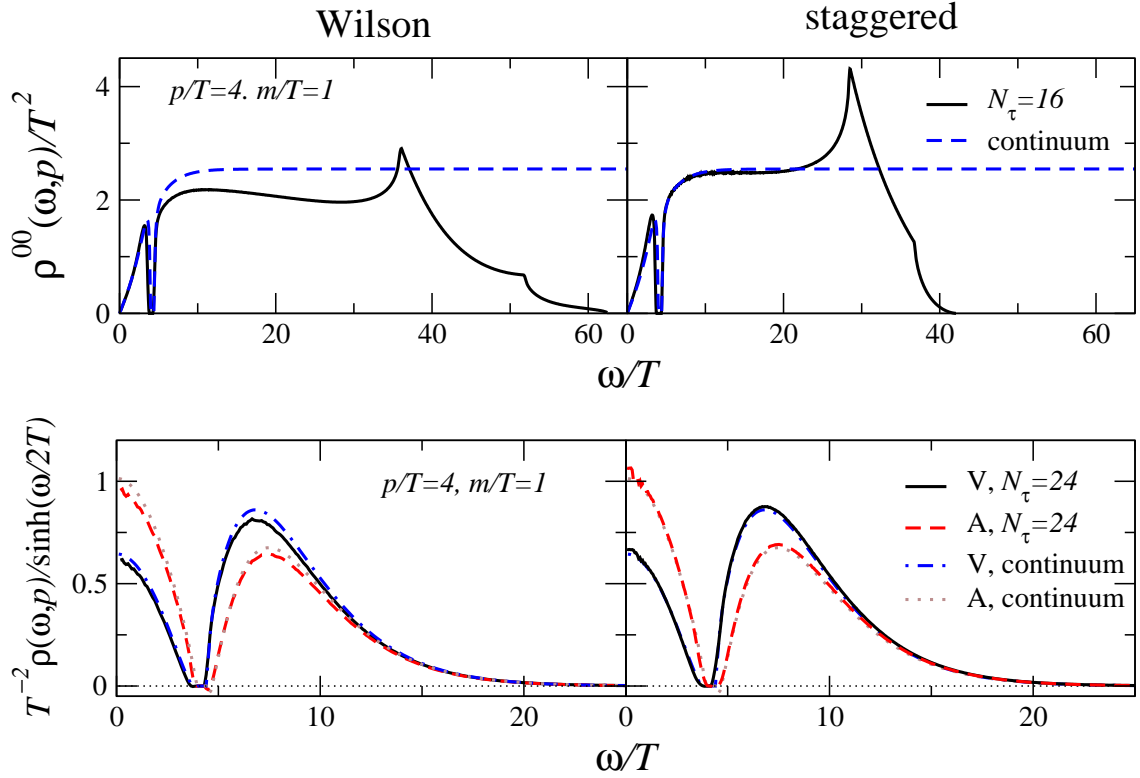

\centerline{\epsfig{figure=plot00.eps,width=15cm}}
\vspace*{0.3cm}
\centerline{\epsfig{figure=plotVAsinh2.eps,width=15cm}}
\caption{Spectral functions $\rho^{00}(\om,\pv)/T^2$ as
a function of $\om/T$ for $N_\tau=16$ (above).
Spectral functions $\rho_{\rm V,A}(\om,\pv)/T^2\sinh(\om/2T)$
as a function of $\om/T$ for $N_\tau=24$ (below).
In both cases $p/T=4, m/T=1$ and $\xi=1$.
}
\label{figplot00}
\end{figure}

The effect of increasing $N_\tau$ is demonstrated in Fig.\
\ref{figplotPSNt} for the pseudoscalar spectral function $\rho_{\rm PS}$
for fixed $\xi=1$ (top) and for the vector spectral function $\rho^{ii}$
for fixed $\xi/N_\tau = aT$ (bottom). As expected from Eqs.\
(\ref{ommaxW}) and (\ref{ommaxS}), $\om_{\rm max}$ increases with
$N_\tau$. In the anisotropic case, a large $N_\tau$ seems to lead to a
better improvement for Wilson than for staggered fermions.
In Fig.\ \ref{figplot00}, we present our results for $\rho^{00}$. As we
emphasized in Eq.\ (\ref{eqasympt2}), due to current conservation this
spectral function does not increase with $\om^2$ for large $\om$, but
instead reaches a constant value $p^2/2\pi$. This can indeed be seen in 
Fig.\ \ref{figplot00}. Due to this behaviour the contribution below the
lightcone is visible in the same plot. In this case it appears that
staggered fermions reproduce the continuum result substantially better
than Wilson fermions.
                                                                                
Finally, we note that the following behaviour of the kernel and spectral 
functions ($\rho^{00}$ and $\rho^{00}_5$ excluded)
\bea
\nn \om\to 0:
&&\hm\;\;\;\;\;\; \rho_H(\om,\pv) \approx \om,
\;\;\;\;\;\;\;
K(\tau,\om) \approx \frac{2T}{\om}, \\
\om\to \infty: &&\hm\;\;\;\;\;\; \rho_H(\om,\pv) \approx \om^2, 
\;\;\;\;\;\;
K(\tau,\om) \approx
e^{-\om\tau} + e^{\om(\tau-1/T)},
\eea
makes it difficult to study spectral functions for both small and
large energies in one plot. This can be circumvented by instead showing 
the integrand at the midpoint $\tau=1/2T$, i.e.,  
\be 
K(1/2T, \om) \rho_H(\om,\pv) =
\frac{\rho_H(\om,\pv)}{\sinh(\om/2T)}, 
\ee 
 which takes a finite value for $\om\to 0$ and vanishes exponentially for 
large $\om$. In Fig.\ \ref{figplot00} we show an example of this for 
$\rho_{\rm V,A}$. Since the region with large $\om$ is exponentially 
suppressed, we note that the lattice artefacts related to the finiteness 
of the Brillioun zone discussed above give an exponentially small 
contribution. We conclude therefore that the euclidean correlator at the 
midpoint $G_H(1/2T,\pv)$ is largely insensitive to these artefacts. We 
also point out that it follows from the analysis in Section \ref{midpoint} 
that the area under the curves are identical: the larger spectral weight 
of $\rho_{\rm A}$ below the lightcone is exactly compensated by the larger 
spectral weight of $\rho_{\rm V}$ above threshold.


\section{Summary}
\label{summary}

We have studied meson spectral functions at nonzero momentum in the 
infinite temperature limit, in the continuum and on the lattice using 
Wilson and staggered fermions. We found that for large values of the 
energy $\om$, lattice spectral functions become sensitive to the effects 
of discretizaton and deviate from the continuum expectation, in agreement 
with the conclusions from Ref.\ \cite{Karsch:2003wy}. For smaller $\om$, 
finite discretization affects predominantly the mismatch between the 
continuum and lattice lightcone, which can be substantial. In the free 
field limit a simple relationship between the euclidean correlators in 
different channels at the midpoint was found.

A qualitative comparison between the results obtained with staggered and 
Wilson fermions suggests that in the low-energy region lattice artefacts 
are less prominent for the staggered formulation. The use of an 
anisotropic lattice, on the other hand, seems to be more beneficial for 
Wilson fermions.


\vspace*{0.5cm}
\noindent
{\bf Acknowledgments.}
 It is a pleasure to thank Simon Hands and Seyong Kim for 
numerous discussions.
 J.M.M.R.\ thanks the Physics Department in Swansea for its hospitality 
during the course of this work.
 G.A.\ is supported by a PPARC Advanced Fellowship.
 J.M.M.R. was supported in part by the Spanish Science Ministry (Grant FPA 
2002-02037) and the University of the Basque Country (Grant 
UPV00172.310-14497/2002).


\end{document}